\documentclass[12pt]{iopart}
\usepackage{amssymb}
\usepackage{txfonts}

\usepackage[dvips]{graphicx}
\usepackage{subfigure}
\usepackage{booktabs}
\usepackage{eurosym}

\usepackage{color}

\begin{document}

\title{Transitions in pedestrian fundamental diagrams of straight corridors
and T-junctions}

\author{J Zhang$^1$, W Klingsch$^1$,  A Schadschneider$^2$ and
A Seyfried$^{3,4}$}

\address{$^1$ Institute for Building Material Technology and Fire
  Safety Science, Bergische Universit\"at Wuppertal,
  Pauluskirchstrasse 11, 42285 Wuppertal, Germany}
\address{$^2$ Institut f\"ur Theoretische Physik, Universit\"at zu
  K\"oln, 50937 K\"oln, Germany}
\address{$^3$ Computer Simulation for Fire Safety and Pedestrian Traffic,
  Bergische Universit\"at Wuppertal, Pauluskirchstrasse 11, 42285
  Wuppertal, Germany}
\address{$^4$ J\"ulich Supercomputing Centre, Forschungszentrum
  J\"ulich GmbH, 52425 J\"ulich, Germany}
\ead{jun.zhang@uni-wuppertal.de, klingsch@uni-wuppertal.de,
as@thp.uni-koeln.de, seyfried@uni-wuppertal.de}

\begin{abstract}
  Many observations of pedestrian dynamics, including various
  self-organization phenomena, have been reproduced successfully by
  different models. But the empirical databases for quantitative
  calibration is still insufficient, e.g. the fundamental diagram as
  one of the most important relationships displays non-negligible
  differences among various studies. To improve this situation,
  experiments in straight corridors and T-junction are performed.
  Four different measurement methods are defined to study their effects
  on the fundamental diagram. It is shown that they have minor influences
  for $\rho <3.5~$m$^{-2}$ but only the Voronoi method is able
  to resolve the fine-structure of the fundamental diagram. This enhanced
  measurement method permits to observe the occurrence of boundary-induced
  phase transition. For corridors of different widths we found that
  the specific flow concept works well for $\rho <3.5~$m$^{-2}$.
  Moreover, we illustrate the discrepancies between the fundamental diagrams
  of a T-junction and a straight corridor.
\end{abstract}

\section{Introduction}

During the last few decades, research on pedestrian and traffic flow
became popular and attracted a lot of attention
\cite{Appert-Rolland2009,Bandini2010,Klingsch2010,Schadschneider2009c,Schadschneider2009,SchadChowNish}.
The investigation of pedestrian motion plays an important role in
guaranteeing the safety of pedestrians in complex buildings or at
mass events. A large number of models have been developed in the
past. Most of them are able to reproduce phenomena of pedestrian
movement qualitatively. Before using a model to predict quantitative
results like the total evacuation time, it needs to be calibrated
thoroughly and quantitatively using empirical data. However, this is
still difficult due to a lack of reliable experimental data. In
addition, the small number of available datasets shows surprisingly
large differences \cite{Schadschneider2009a,Seyfried2009}.

In recent years, several well-controlled pedestrian experiments
\cite{Hoogendoorn2005,Kretz2006a,Kretz2006,Moussaid2009,Liu2009} and
field studies \cite{Johansson2009a,Johansson2008,Young1999} have
been performed. One of the most important characteristics of
pedestrian dynamics is the fundamental diagram which states the
relationship between pedestrian flow and density. Several
researchers, in particular Fruin and Pauls \cite{Fruin1971},
Predtechenskii and Milinskii \cite{Predtechenskii1978}, Weidmann
\cite{Weidmann1993}, Helbing et al. \cite{Helbing2007} have
collected information about the relation of occupants density and
velocity. But there exists considerable disagreement among these
data. In the comparison performed in \cite{Seyfried2010} the density
$\rho_0$, where the velocity approaches zero due to overcrowding,
ranges from 3.8~m$^{-2}$ to 10~m$^{-2}$, while the density $\rho_c$
where the flow reaches its maximum ranges from 1.75~m$^{-2}$ to
7~m$^{-2}$. Several explanations for these discrepancies have been
proposed, including cultural factors \cite{Chattaraj2009},
differences between unidirectional and multidirectional flow
\cite{Navin1969,Pushkarev1975}.

For single-file movements it was found that even the measurement
method has a large influence on the fundamental diagram and could be
responsible for the observed deviations \cite{Seyfried2010}.  For
identical facilities (e.g., corridors, stairs, doors) with different
width it is usually assumed that the fundamental diagrams are unified
in a single diagram for specific flow $J_s$. The study of Hankin et al.\
\cite{Hankin1958} in the London subway shows that above a certain minimum
of about four feet the maximum flow in subway stations is directly
proportional to the width of the corridor.  Besides, it is still not
clear whether or not the fundamental diagrams for other scenarios like
bottlenecks or T-junctions are the same.

Facing such questions, a series of well-controlled laboratory
experiments are carried out. The goals of this study are to improve
the database related to pedestrian dynamics, to determine the influence
of measurement methods on the fundamental diagram, and to check
whether or not the fundamental diagram for different types of
facilities can be unified in a single diagram.

In section $2$, the experiment setup will be briefly described. The
measurement methods used in this paper are introduced and defined in
section $3$. The main results of the work are in section $4$.
Finally, we make some concluding remarks in section $5$.

\section{Experiment Setup}

The experiments were performed in hall~2 of the fairground
D\"usseldorf (Germany) in May~2009. They are part of the Hermes project
\cite{Hermes} in which the data resulting from the experiments will be
used to calibrate and test pedestrian movement models. The experiments
were conducted with up to 350 participants.  They were composed mostly
of students and each of them was paid 50~{\euro} per day. The mean age
and height of the participants were $25 \pm 5.7$ years and $1.76 \pm
0.09$~m, respectively. The free velocity $v_0= 1.55 \pm 0.18$~m/s was
obtained by measuring 42 participants' free movement.

\begin{figure}
\centering\subfigure[Sketch of experiment at straight corridor]{
\includegraphics[scale=2.0]{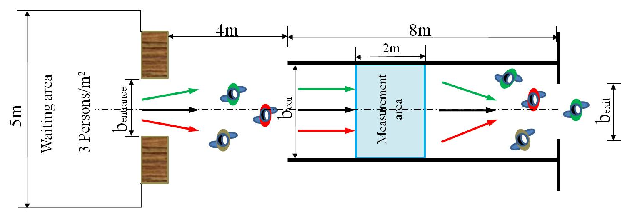}}
\subfigure[Snapshots from experiment $C$-180-180-070]{
\includegraphics[scale=1.9]{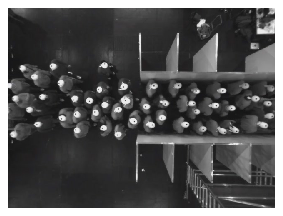}\qquad
\includegraphics[scale=1.9]{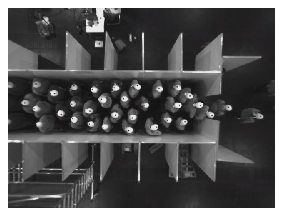}}
\subfigure[Sketch of experiment at T-junction]{
\includegraphics[scale=2.0]{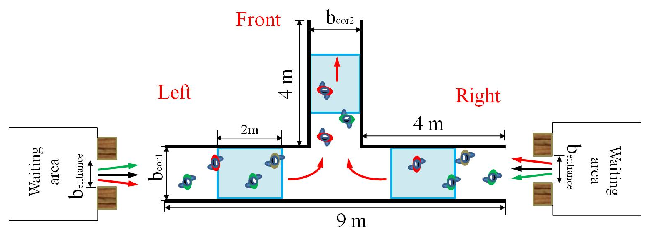}}
\subfigure[Snapshots from experiment $T$-240-120-240]{
\includegraphics[scale=2.25,angle=180]{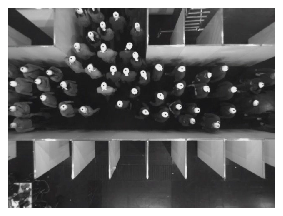}\qquad
\includegraphics[scale=1.7,angle=-90]{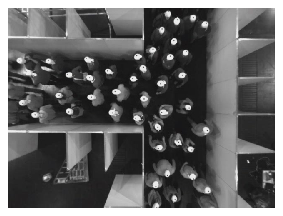}}
\caption{\label{fig1} Sketch of the experimental setup ((a) and
(c)) and snapshots of the experiment from the two cameras
respectively ((b) and (d)). The shaded regions in (a) and (c) are
the chosen measurement areas with 2~m in length.}
\end{figure}

Figure~\ref{fig1} shows the sketches of the setups and some
snapshots during the experiments. Two types of geometries, straight
corridor ($C$) and T-junction ($T$), were used in the experiment. 28
runs (see Table \ref{table2}) were performed in straight corridors
with widths of 1.8~m, 2.4~m and 3.0~m respectively. 7 runs for the
T-junction with corridor width of 2.4~m were carried out (see
Table~\ref{table3}). To regulate the pedestrian density in the
corridor, the width of the entrance $b_{\rm entrance}$ and the exit
$b_{\rm exit}$ was changed in each run. For details, see
Figure~\ref{fig1}, Table~\ref{table2} and \ref{table3}. At the
beginning, the participants were held within a waiting area. Equal
densities for different runs were arranged by partitioning the
waiting area and counting the number of people in the parts.
Standing at the waiting area, they pass through a 4~m passage into
the corridor. The passage was used as a buffer to minimize the
effect of the entrance. In this way, the pedestrian flow in the
corridor was nearly homogeneous over its entire width. When a
pedestrian leaves through an exit, he or she returns to the waiting
area for the next run.

\begin{figure}
\centering\subfigure[$C$-180-180-070]{
\includegraphics[scale=1.0]{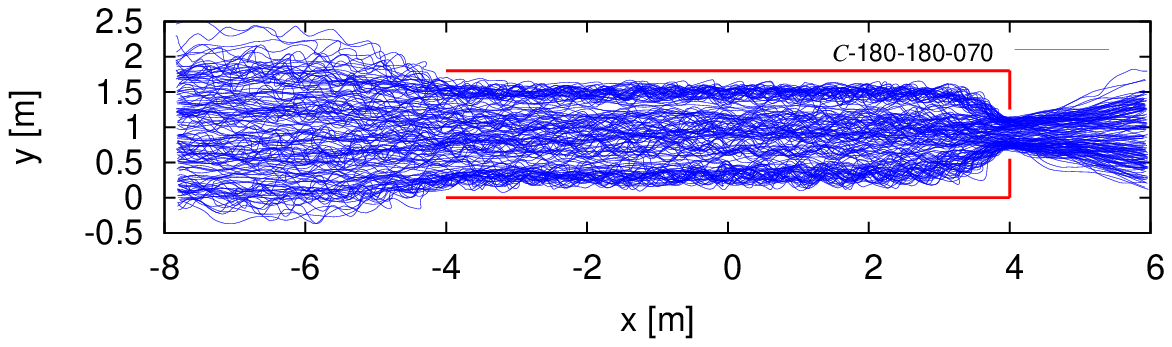}}
\subfigure[$T$-240-100-240]{
\includegraphics[scale=1.0]{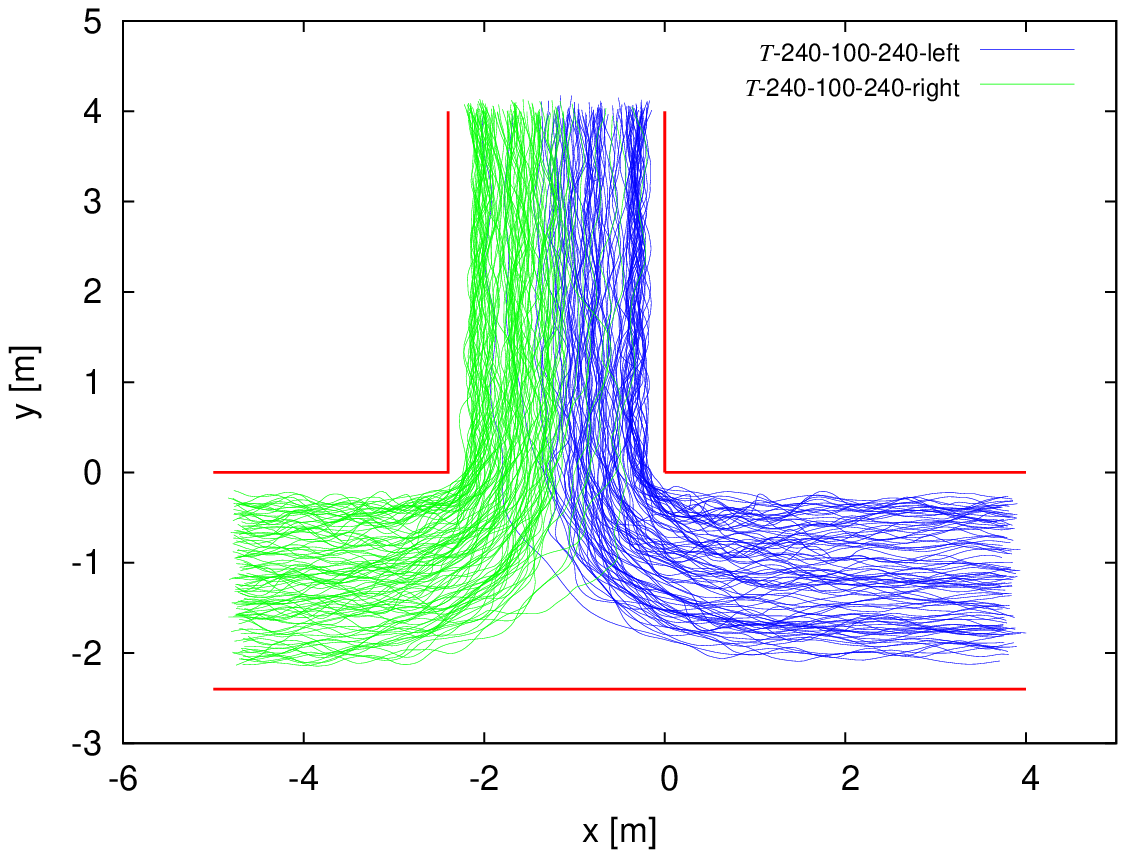}}
\caption{\label{fig2} The trajectories for all pedestrians in the
experiment. The data are extracted from the video recording by the
free software {\em PeTrack} \cite{Boltes2010}. }
\end{figure}

The experiments were recorded by two cameras mounted at the rack of
the ceiling of the hall. To cover the complete region, the left and
the right part of the corridor were recorded by the two cameras
separately. The pedestrian trajectories were automatically extracted
from video recordings using the software {\em PeTrack}
\cite{Boltes2010}. Finally, the trajectory data from the two cameras
were corrected manually and combined automatically. The frame rate
of the trajectory data corresponds to 16~fps. Figure~\ref{fig2}
shows the trajectories of the head of each pedestrian in two runs of
the experiment. From these trajectories, pedestrian characteristics
including flow, density and velocity are determined.

\section{Measurement Methods}

For vehicular traffic it is well known that different measurement
methods lead to different fundamental diagrams
\cite{Leutzbach1988,Kerner2004}. The results presented in
\cite{Seyfried2010} using pedestrian trajectories of single file
movement, have also shown how large variations induced by different
measurement methods could be. In previous studies of pedestrian
streams, different measurement methods were used limiting the
comparability of the data. E.G. Helbing et al. proposed a Gaussian,
distance-dependent weight function \cite{Helbing2007} to measure the
local density and local velocity. Predtechenskii and Milinskii
\cite{Predtechenskii1978} used a dimensionless definition to
consider different body sizes and Fruin introduced the "Pedestrian
Area Module" \cite{Fruin1971}. All of these definitions have their
advantages and disadvantages. To analyze all of them goes beyond the
scope of this paper. To enable a detailed analysis, we study the
influence of several measurement methods on the fundamental diagram
and analyze which methods lead to the smallest fluctuations.

In this study four measurement methods were used to calculate the
basic quantities: flow, density and velocity. Some terminologies used
here are taken from \cite{Seyfried2010} and \cite{Steffen2010a}.

\begin{figure}
\centering\includegraphics[scale=1.8]{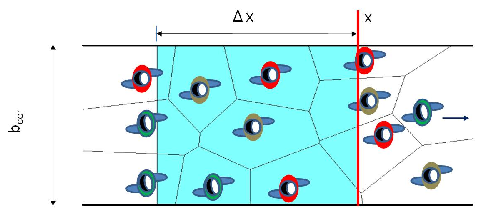}
\caption{\label{fig3} Illustration of different measurement methods.
$Method$ $A$ is a kind of local measurement at cross-section with
position $x$ averaged over a time interval $\Delta t$, while
$Methods$ $B-D$ measure at a certain time and average the results
over space $\Delta x$. Note that for $Method$ $D$, the Voronoi
diagrams are generated according to the spatial distributions of
pedestrians frame by frame.}
\end{figure}

\subsection{Method A}

Method A calculates mean value of flow and density over time. A
reference location $x$ in the corridor is taken and studied over a
fixed period of time $\Delta {t}$ (as shown in Figure~\ref{fig3}).  We
refer to this average by $\langle \rangle_{\Delta t}$. Using this
method we can obtain the pedestrian flow $J$ and the velocity $v_i$ of
each pedestrian passing $x$ directly. Thus, the flow over time
$\langle J \rangle_{\Delta t}$ and the time mean velocity $\langle v
\rangle_{\Delta t}$ can be calculated as
\begin{equation}\label{eq1}
\langle J \rangle_{\Delta t}=\frac{N_{\Delta t}}{t_{N_{\Delta
t}}}\qquad {\rm and} \qquad \langle v \rangle_{\Delta
t}=\frac{1}{N_{\Delta t}}\sum_{i=1}^{N_{\Delta t}} v_i(t)
\end{equation}
where $N_{\Delta t}$ is the number of persons passing the location
$x$ during the time interval $\Delta t$. $t_{N_{\Delta t}}$ is the
time between the first and the last of the $N_{\Delta t}$
pedestrians. Thus $t_{N_{\Delta t}}$ is the actual time
that the $N_{\Delta t}$ pedestrians used for passing the location.
It can be different from $\Delta t$. The time mean velocity
$\langle v \rangle_{\Delta t}$ is defined as the mean value of the
instantaneous velocities $v_i(t)$ of the $N_{\Delta t}$ persons
according to equation (\ref{eq6}). We calculate $v_i(t)$ by use of
the displacement of pedestrian $i$ in a small time interval $\Delta
t^\prime$ around $t$:
\begin{equation}\label{eq6}
v_i(t)=\frac{{x_i}(t+\Delta t^\prime/2)-{x_i}(t-\Delta
t^\prime/2))}{\Delta t^\prime}\,
\end{equation}

\subsection{Method B}

The second method measures the mean value of velocity and density
over space and time. The spatial mean velocity and density are
calculated by taking a segment $\Delta x$ in the corridor as the
measurement area. The velocity $\langle v \rangle_i$ of each person
is defined as the length $\Delta x$ of the measurement area divided
by the time he or she needs to cross the area (see equation~(\ref{eq2})),
\begin{equation}\label{eq2}
\langle v \rangle_i=\frac{\Delta x}{t_{\rm out}-t_{\rm in}}\,
\end{equation}
where $t_{\rm in}$ and $t_{\rm out}$ are the times a person enters
and exits the measurement area, respectively.
The density $\rho_i$ for each person is calculated with equation (\ref{eq3}):
\begin{equation}\label{eq3}
\langle \rho \rangle_i=\frac{1}{t_{\rm out}-t_{\rm
in}}\cdot\int_{t_{\rm in}}^{t_{\rm out}} \frac{N^\prime(t)}{b_{\rm
cor}\cdot\Delta x}dt\,
\end{equation}
$b_{\rm cor}$ is the width of the measurement area while $N^\prime(t)$
is the number of person in this area at a time $t$.

\subsection{Method C}

The third measurement method is the classical method. The density
$\langle \rho \rangle_{\Delta x}$ is defined as the number of
pedestrians divided by the area of the measurement section:
\begin{equation}\label{eq4}
\langle \rho \rangle_{\Delta x}=\frac{N}{b_{\rm cor}\cdot\Delta x}\,
\end{equation}
The spatial mean velocity is the average of the instantaneous
velocities $v_i(t)$ for all pedestrians in the measurement area at
time $t$:
\begin{equation}\label{eq5}
\langle v \rangle_{\Delta x}=\frac{1}{N}\sum_{i=1}^{N}{v_i(t)}\,
\end{equation}

\subsection{Method D}

This method is based on the use of Voronoi diagrams
\cite{Voronoi1908} which are a special kind of decomposition of a
metric space determined by distances to a specified discrete set of
objects in the space. At any time the positions of the pedestrians
can be represented as a set of points, from which the Voronoi
diagram (see Figure \ref{fig3}) can be generated. The Voronoi cell
area, $A_i$, for each person $i$ can be obtained. Then, the density
and velocity distribution of the space $\rho_{xy}$  and $ v_{xy}$
can be defined as
\begin{equation}\label{eq9}
\rho_{xy} = 1/A_i \quad and \quad v_{xy}={v_i(t)}\qquad \mbox{if
$(x,y)\in A_i$}\,
\end{equation}
where $v_i(t)$ is the instantaneous velocity of each person, see
equation~(\ref{eq6}). The Voronoi density and velocity for the
measurement area is defined as \cite{Steffen2010a}
\begin{equation}\label{eq7}
\langle \rho \rangle_v=\frac{\iint{\rho_{xy}dxdy}}{b_{\rm
cor}\cdot\Delta x}\,
\end{equation}
\begin{equation}\label{eq8}
\langle v \rangle_v=\frac{\iint{v_{xy}dxdy}}{b_{\rm cor}\cdot\Delta
x}\,
\end{equation}

\section{Results}

We calculate the fundamental diagram for the straight corridor
experiments using the methods introduced in the last section. To
facilitate a comparison among these four methods, we use the
hydrodynamic flow equation $J=\rho vb$.

\subsection{Influence of the measurement method}

\begin{figure}
\subfigure[Density]{
\centering\includegraphics[scale=0.8]{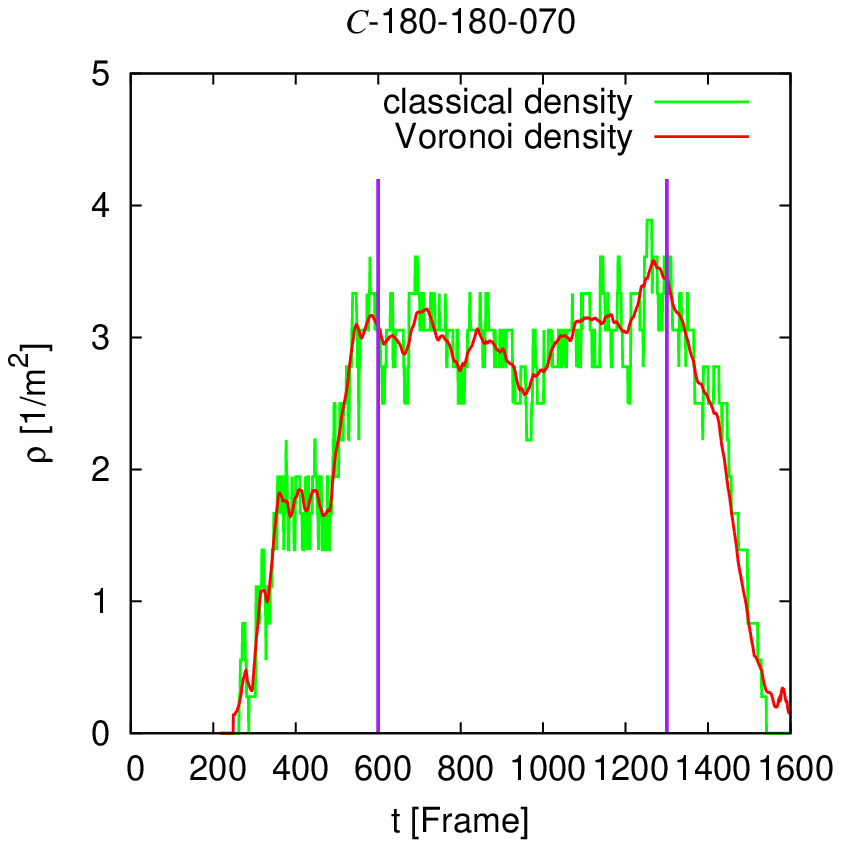}}
\subfigure[Velocity]{
\includegraphics[scale=0.8]{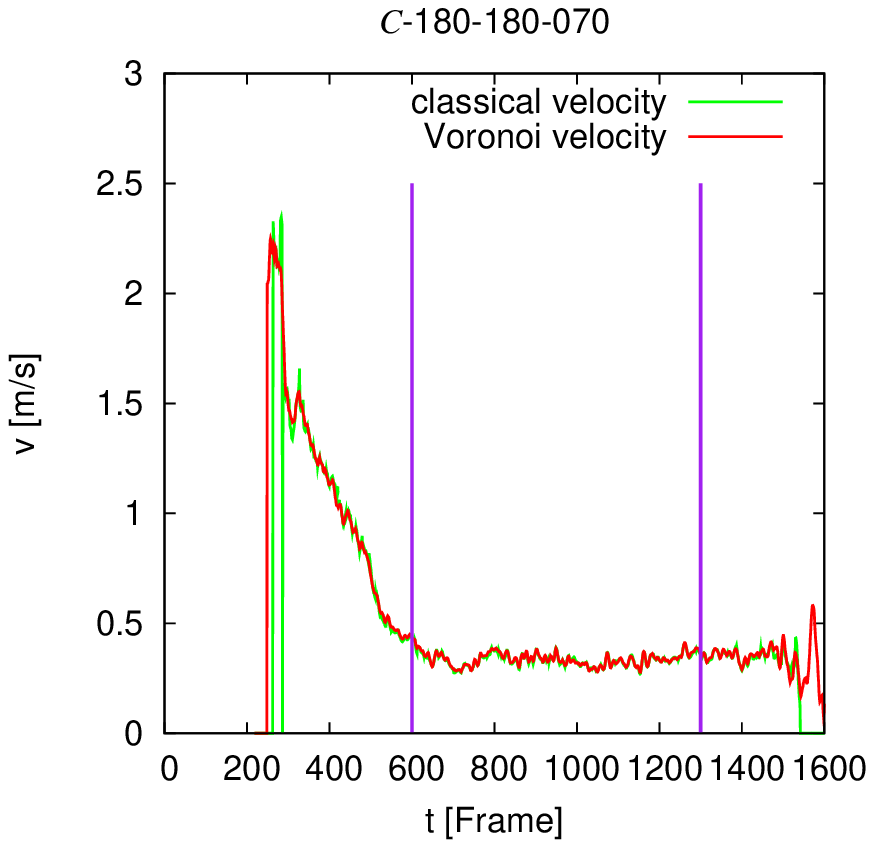}}
\caption{\label{fig4} Time series of density and velocity using
Method $C$ and $D$. The two vertical lines indicate the beginning
and the end of stationary states.}
\end{figure}

For $Method$ $A$ we choose the time interval $\Delta t=10$~s,
$\Delta t^\prime=0.625$~s (corresponding to 10 frames) and the
measurement position at $x=0$ (see Figure~\ref{fig2}). For the other
three methods a rectangle with length 2~m from $x=-2$~m to $x=0$ and
width of the corridor is chosen as the measurement area. We
calculate the densities and velocities every frame with a frame rate
of 16~fps. All data below are obtained from the trajectories. To
determine the fundamental diagram only data at stationary states,
which were selected manually by analyzing the time series density
and velocity (see Figure~\ref{fig4}), were used. For $Method$ $D$ we
use one frame per second to decrease the number of data points and
to represent the data more clearly.

\begin{figure}
\subfigure[$Method$ $A$]{
\centering\includegraphics[scale=0.9]{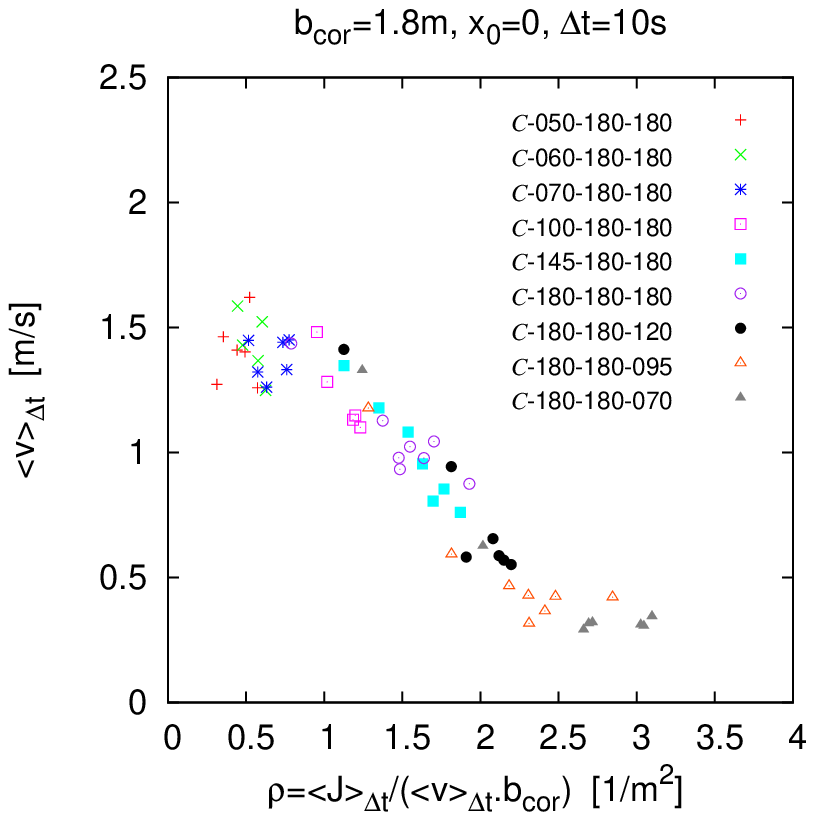}}
\subfigure[$Method$ $B$]{
\includegraphics[scale=0.9]{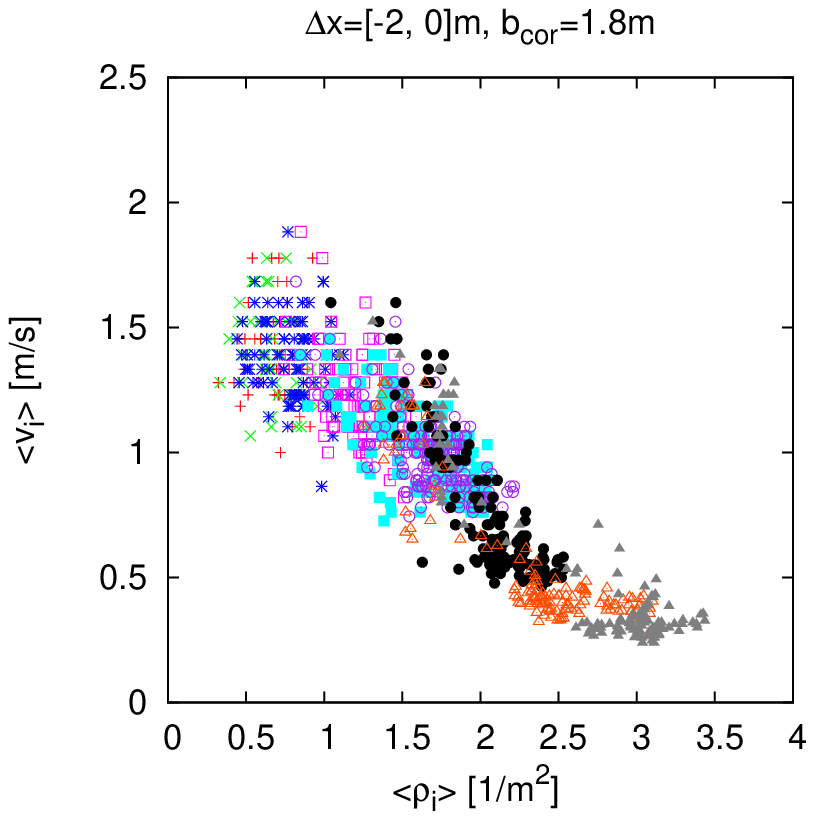}}
\subfigure[$Method$ $C$]{
\includegraphics[scale=0.9]{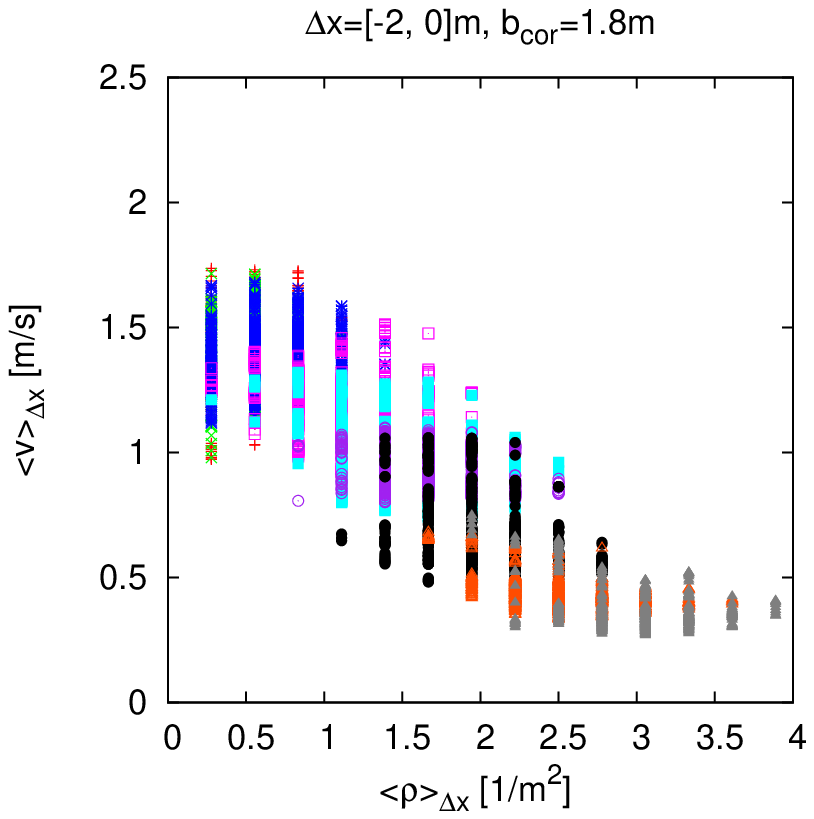}}
\subfigure[$Method$ $D$]{
\includegraphics[scale=0.9]{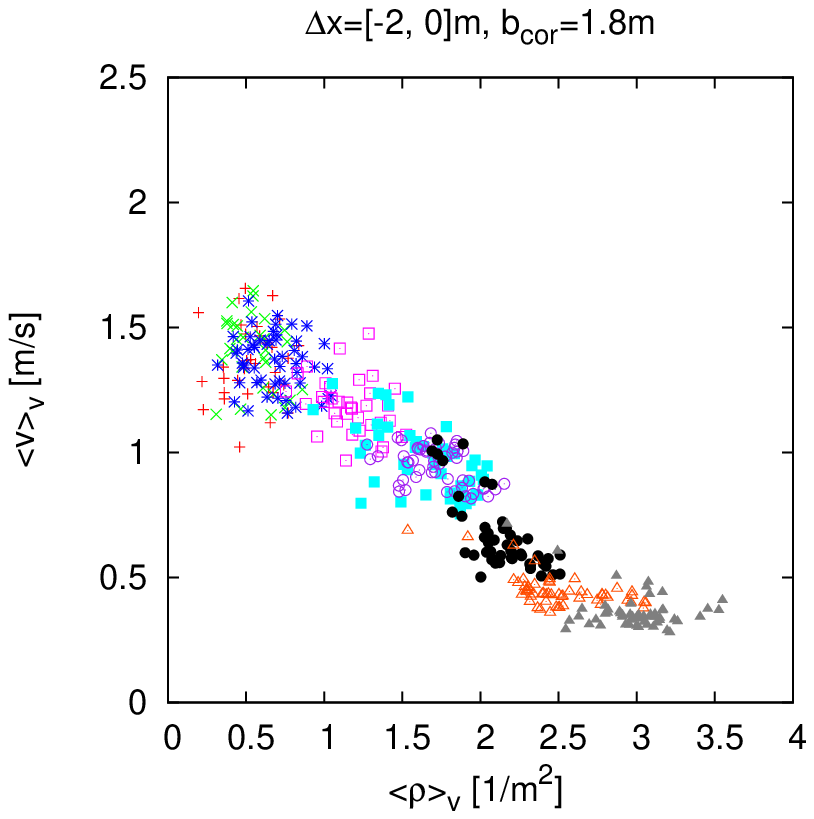}}
\caption{\label{fig5} The fundamental diagrams, the relationship
between density and velocity, measured on the same set of
trajectories but with different methods. Except for the density in (a),
which is calculated using $\rho=J/(b\cdot\Delta x)$, all data are
determined directly from the trajectories. The legends in
(b), (c) and (d) are the same as in (a).}
\end{figure}

Figure~\ref{fig5} shows the relationship between density and velocity
obtained from different methods. Using $Method$ $A$ the flow and mean
velocity can be obtained directly. To get the relationship between
density and flow, the equation $\rho=\langle J\rangle_{\Delta
  t}/(\langle v\rangle_{\Delta t}\cdot b_{\rm cor})$ was adopted to
calculate the density. For the $Method$ $B$, $C$ and $D$ the mean
density and velocity can be obtained directly since they are mean
values over space. There exists a similar trend of the fundamental
diagram obtained using the different methods. However, their
influence on the scatter of the results is obvious.
Table~\ref{table1} shows the standard deviation of velocities in
certain density intervals for different measurement methods.
Compared to the other approaches, the fluctuations in $Method$ $B$
and $C$ are larger.

Another criterion for the quality of the methods is the resolution
in time and space. Even if $Method$~$A$ provides a smaller standard
deviation than $Method$~$D$, the low resolution in time smears the
transition at $\rho=2 m^{-2}$ clearly visible in
Figure~\ref{fig5}(d). The density in $Method$~$C$ has a strong
dependence on the size of the measurement area $A_m=b_{\rm
cor}\cdot\Delta x$. The interval between two density values is
$1/A_m$, which indicates that the measurement area should not be too
small using this method. But large areas can not provide a detailed
spatial resolution. $Method$ $D$ can reduce the density and velocity
scatter \cite{Steffen2010a}. The reduced fluctuation of $Method$ $D$
is combined with a good resolution in time and space, which reveal a
phenomenon that is not observable with $Method$ $A$, $B$ and $C$. In
Figure~\ref{fig5}(d) it seems that there is a discontinuity of the
fundamental diagram when the density is about 2~m$^{-2}$. This will
be analyzed in detail in section~\ref{subsec-disont}.

\begin{figure}
\subfigure[$Method$ $A$]{
\centering\includegraphics[scale=0.9]{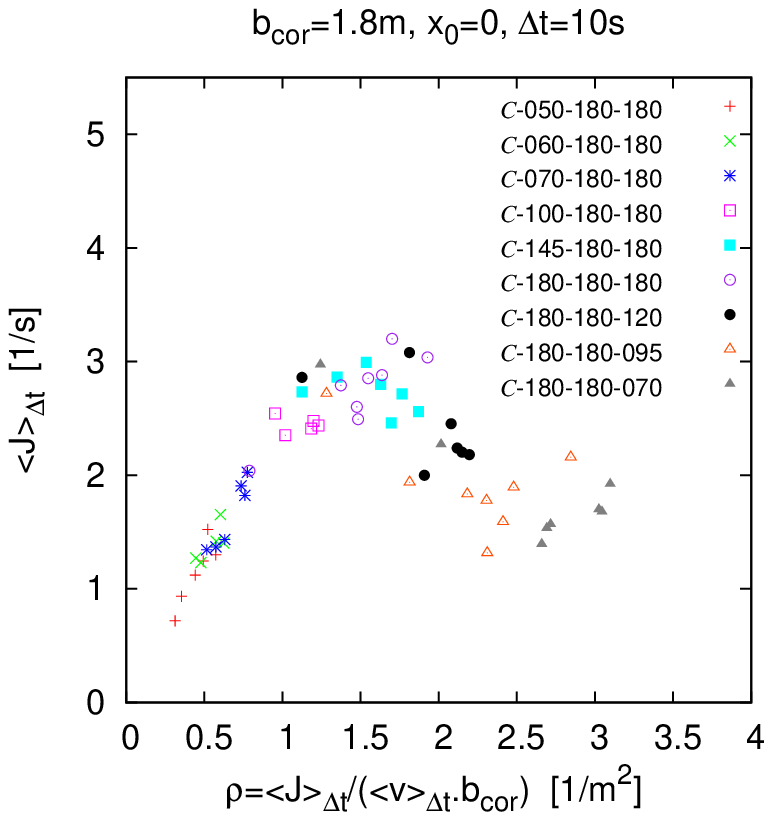}}
\subfigure[$Method$ $B$]{
\includegraphics[scale=0.9]{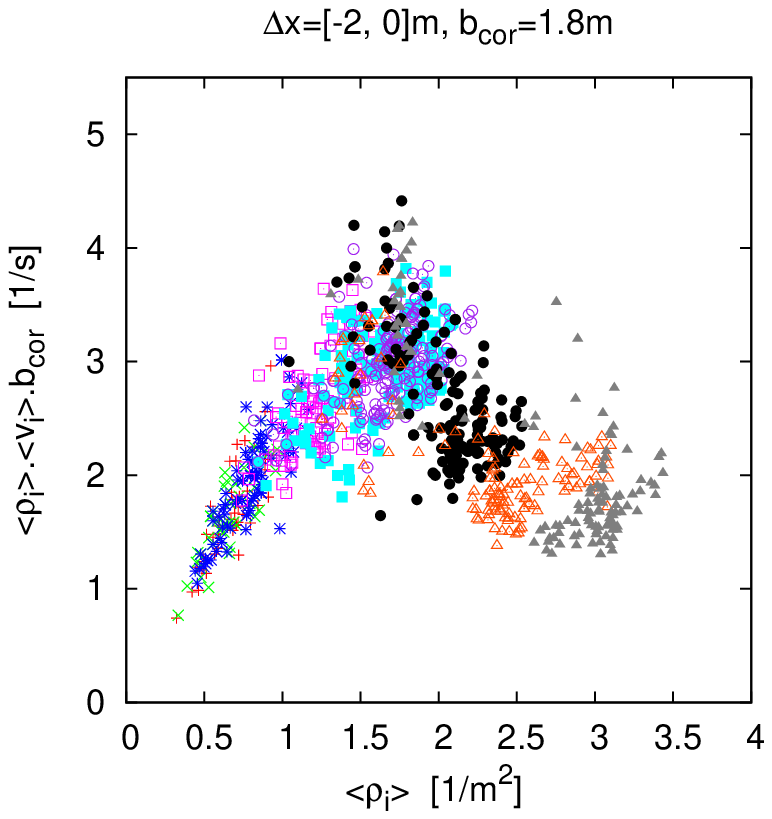}}
\subfigure[$Method$ $C$]{
\includegraphics[scale=0.9]{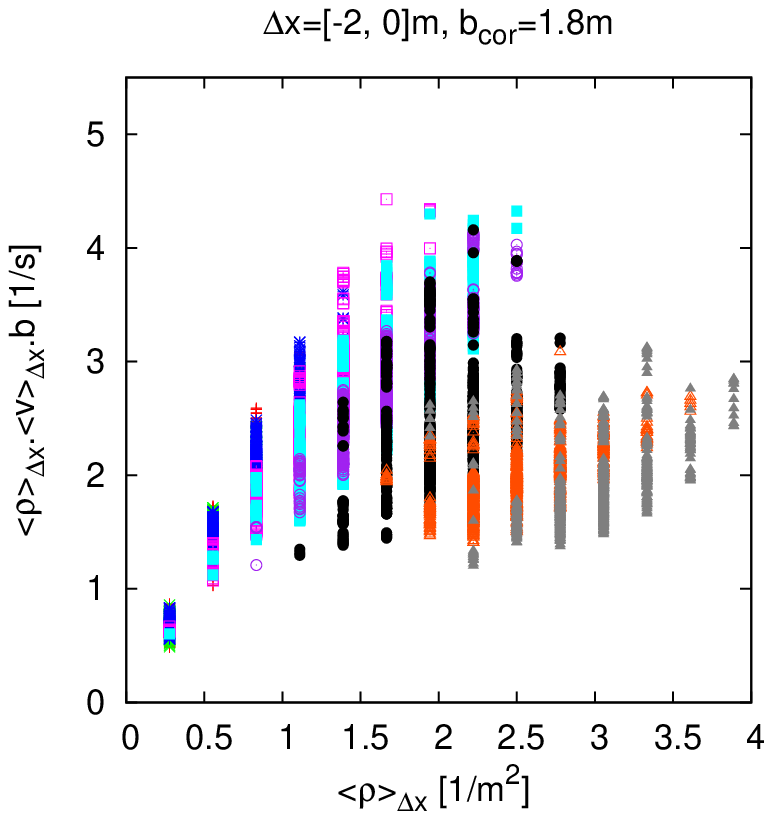}}
\subfigure[$Method$ $D$]{
\includegraphics[scale=0.9]{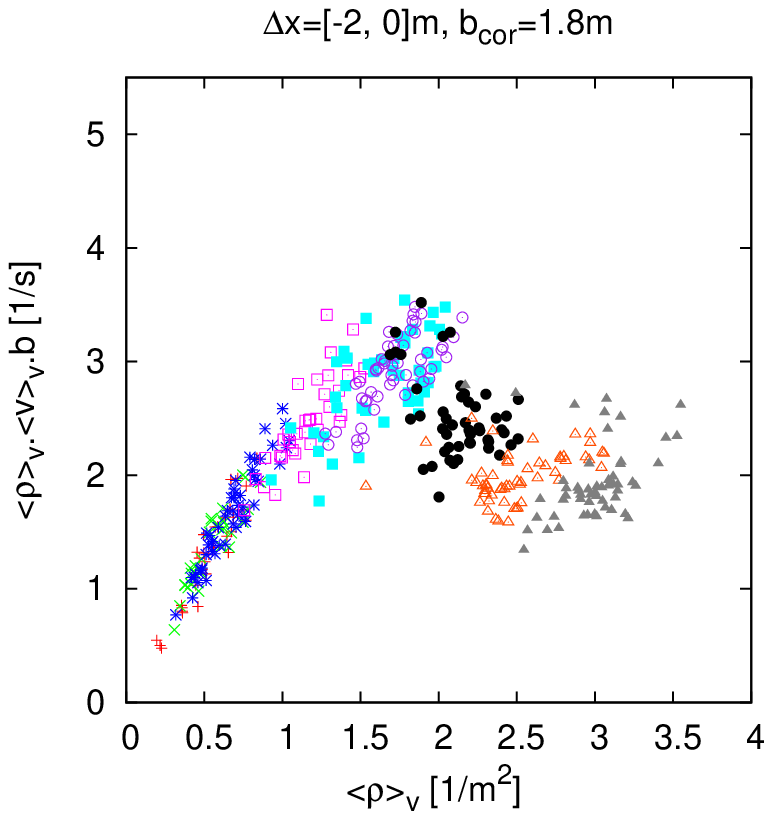}}
\caption{\label{fig6} The fundamental diagrams, the relationship
  between density and flow, measured at the same set of trajectories
  but with different methods. The density in (a) is calculated
  indirectly using $\rho=J/(b\cdot\Delta x)$, while the flows in (b),
  (c) and (d) are obtained by adopting the equation $J=\rho vb$. The
  legends in (b), (c) and (d) are the same as in (a).}
\end{figure}

Figure~\ref{fig6} shows the relationship between the density and
flow obtained from different methods. The pedestrian flow shows
small fluctuations at low densities and high fluctuations at high
densities. The fluctuations for $Method$ $A$ and $Method$ $D$ are
smaller than that for other methods. However, there is a major
difference between the results. While the fundamental diagrams
obtained using $Method$ $A$ and $Method$ $C$ are smooth, fundamental
diagrams obtained with $Method$ $B$ and $Method$ $D$ show a clear
discontinuity at a density of about 2~m$^{-2}$. The average over a
time interval of $Method$ $A$ and the large scatter of $Method$ $C$
blur this discontinuity. In disagreement with the results in
\cite{Seyfried2010}, no marked differences occur among the
fundamental diagrams produced by different methods (see
Figure~\ref{fig5}). In \cite{Seyfried2010}, single-file movement in
a corridor with periodic boundary was studied. In that experiment,
distinct stop waves occured at high densities and lead to large
inhomogeneities in the trajectories. Possibly the differences of
different methods will be larger in the cases where stop waves occur
or when the characteristic of the pedestrian flow is not laminar.

\begin{table}[htbp]
 \centering\caption{\label{table1}Standard deviation of velocities in certain density interval for different methods}
 \lineup
 \begin{tabular}{cccccc}
  \toprule
  Density interval [m$^{-2}$] & $Method$ $A$ [m/s] & $Method$ $B$ [m/s] & $Method$ $C$ [m/s] & $Method$ $D$ [m/s] \\
  \midrule
  $\rho \in [0.8,1.2]$ & 0.119 & 0.169 & 0.175 & 0.120 \\
  $\rho \in [1.6,2.0]$ & 0.086 & 0.144 & 0.175 & 0.111 \\
  \bottomrule
 \end{tabular}
\end{table}

\subsection{Influence of the corridor width}

From the above analysis it can be concluded that there is no large
influence of different methods on the results for the density region
without the stop wave phenomenon. We have shown that the
results with the smallest fluctuations are provided by $Method$ $A$
and $D$. Thus we analyze the other unidirectional experiments with
corridor width 2.4~m and 3.0~m using $Method$~$A$ and $Method$~$D$.

\begin{figure}
\subfigure[Density-velocity using $Method$ $A$]{
\centering\includegraphics[scale=0.9]{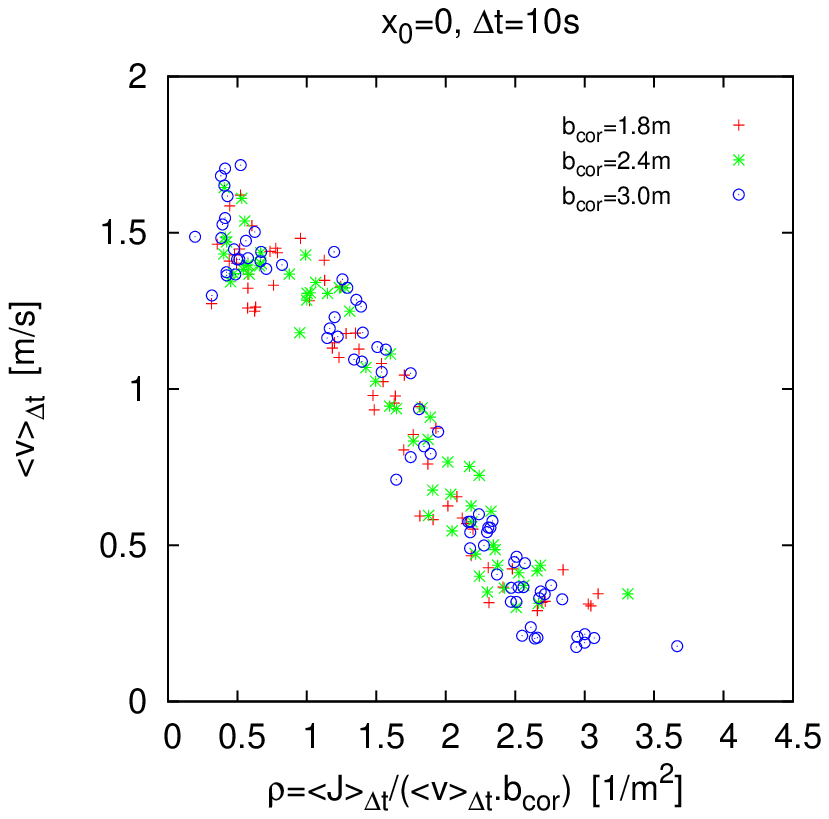}}
\subfigure[Density-velocity using $Method$ $D$]{
\includegraphics[scale=0.9]{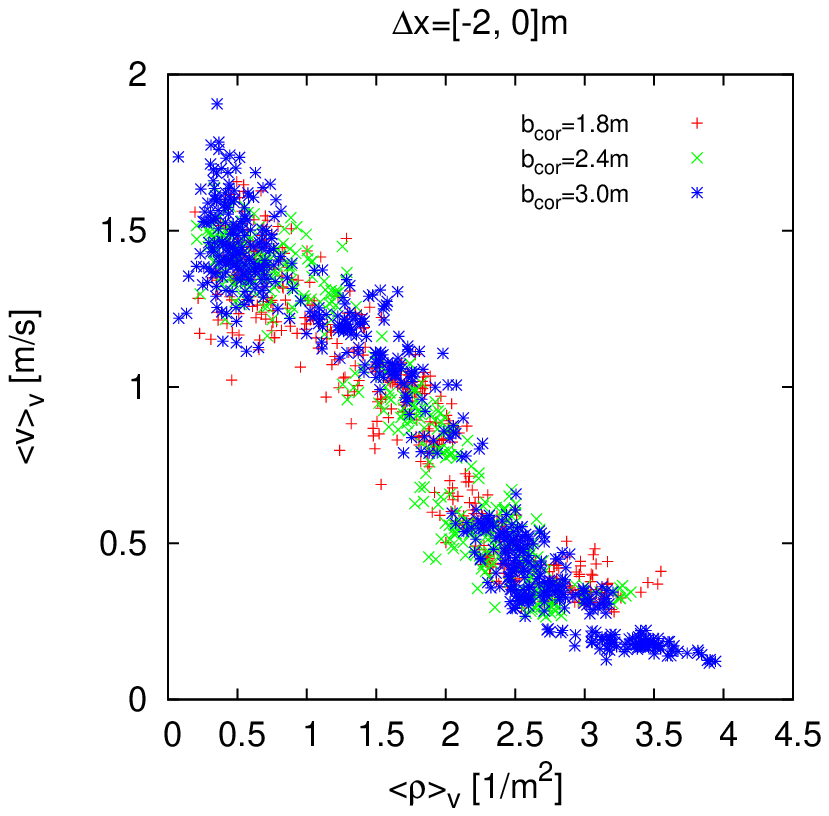}}
\subfigure[Density-specific flow using $Method$ $A$]{
\includegraphics[scale=0.9]{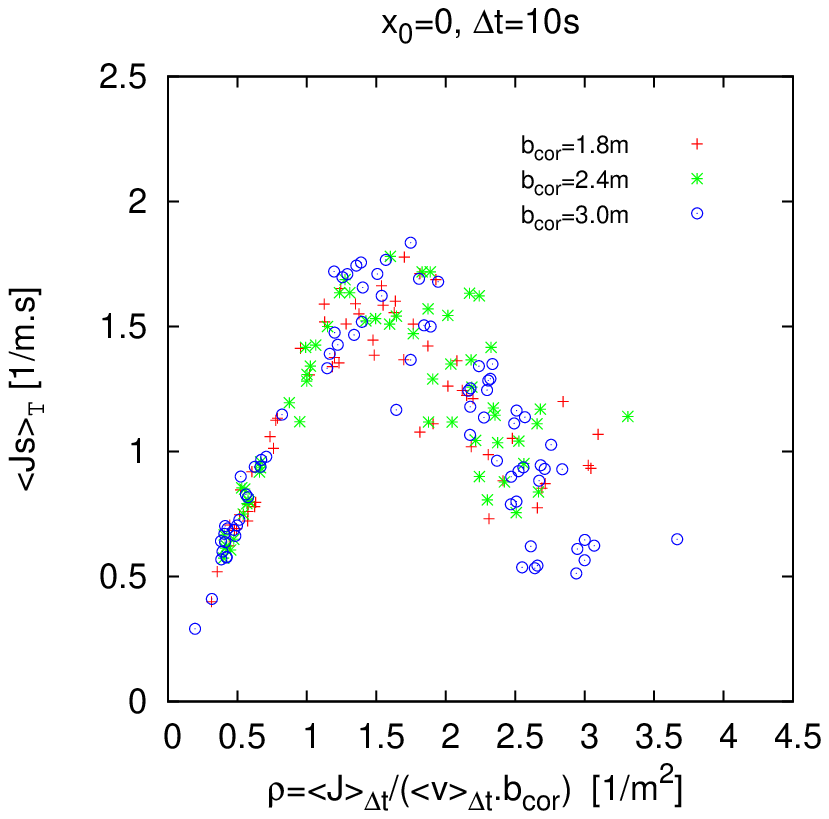}}
\subfigure[Density-specific flow using $Method$ $D$]{
\includegraphics[scale=0.9]{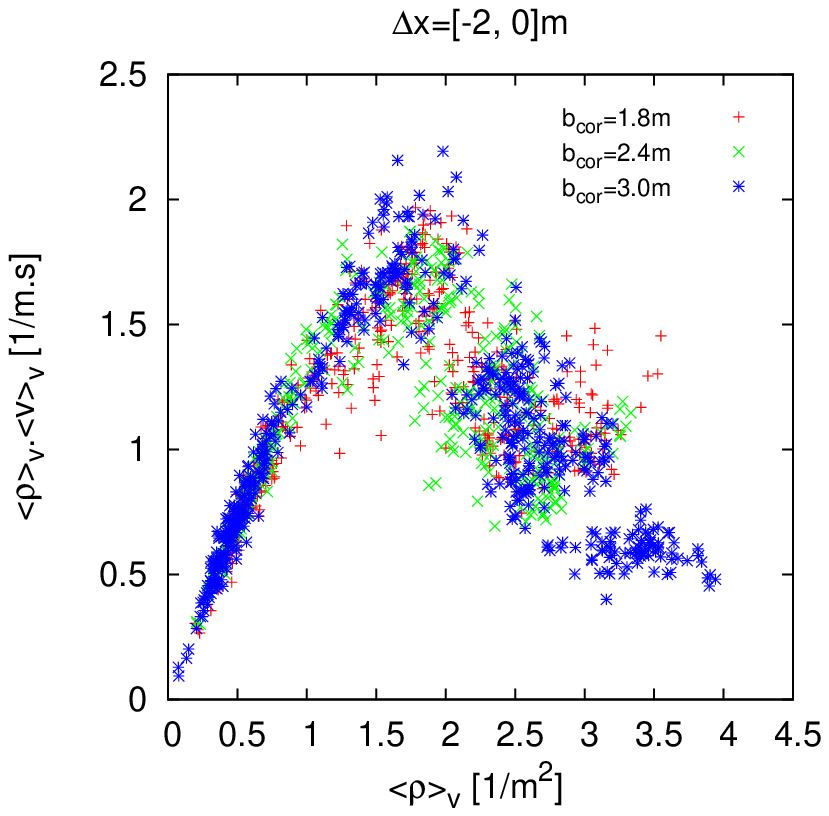}}
\caption{\label{fig7} Comparison of the fundamental diagram in
different corridor widths. }
\end{figure}

Figure~\ref{fig7} shows the relationship between density, velocity and
flow using these two methods. The fundamental diagram of the same
type of corridor but with different widths are compared. The
fundamental diagrams for these three widths agree well for both
methods. This result is in conformance with Hankin's findings
\cite{Hankin1958}. He found that above a certain minimum of about
4~ft (about 1.22~m) the maximum flow in subways is directly
proportional to the width of the corridor. Our results agree with
the assumption that the specific flow $J_s = J/b$ is independent of
the width of the facility. However, it is possible that for small
corridors or very high densities $J_s$ becomes dependent on $b$.

\subsection{Discontinuous trend of the fundamental diagram}
\label{subsec-disont}

In Figure~\ref{fig5}(b) and \ref{fig5}(d) a discontinuity occurs at
$\rho \thickapprox 2$~m$^{-2}$, separating the function $J_s(\rho)$ in
a region $\rho < 2$~m$^{-2}$ with negative curvature and a region
$\rho > 2$~m$^{-2}$ with positive curvature. Moreover, for
$Method$~$D$ a gap occurs around $v = 0.7$~m/s. This transition is
also found in the experiments with $b_{\rm cor} = 2.4$~m and $b_{\rm
  cor} = 3.0$~m. The fundamental diagram changes qualitatively when
the width $b_{\rm exit}$ of the exit is modified. The modification
of the exit width was necessary to achieve high densities. However,
it seems that this slight change in the experimental setup causes a
significant change in the flow-density relation. This point becomes
obvious in the velocity-density relation, especially in
Figure~\ref{fig5}(d) which is obtained from $Method$~$D$.  Although
the measurement area in the corridor is 4~m away from the exit, the
influence of the change of the exit on the fundamental diagram is
sensitive. The decreasing of the exit width limits the outflow of
pedestrians and leads to a discontinuity in the fundamental diagram.
This can be interpreted in terms of the well-established theory of
boundary induced phase transitions, see
section~\ref{subsec-boundary}.

\subsection{Interpretation in terms of boundary-induced phase transitions}
\label{subsec-boundary}

\begin{figure}
\begin{center}
\includegraphics[scale=0.5]{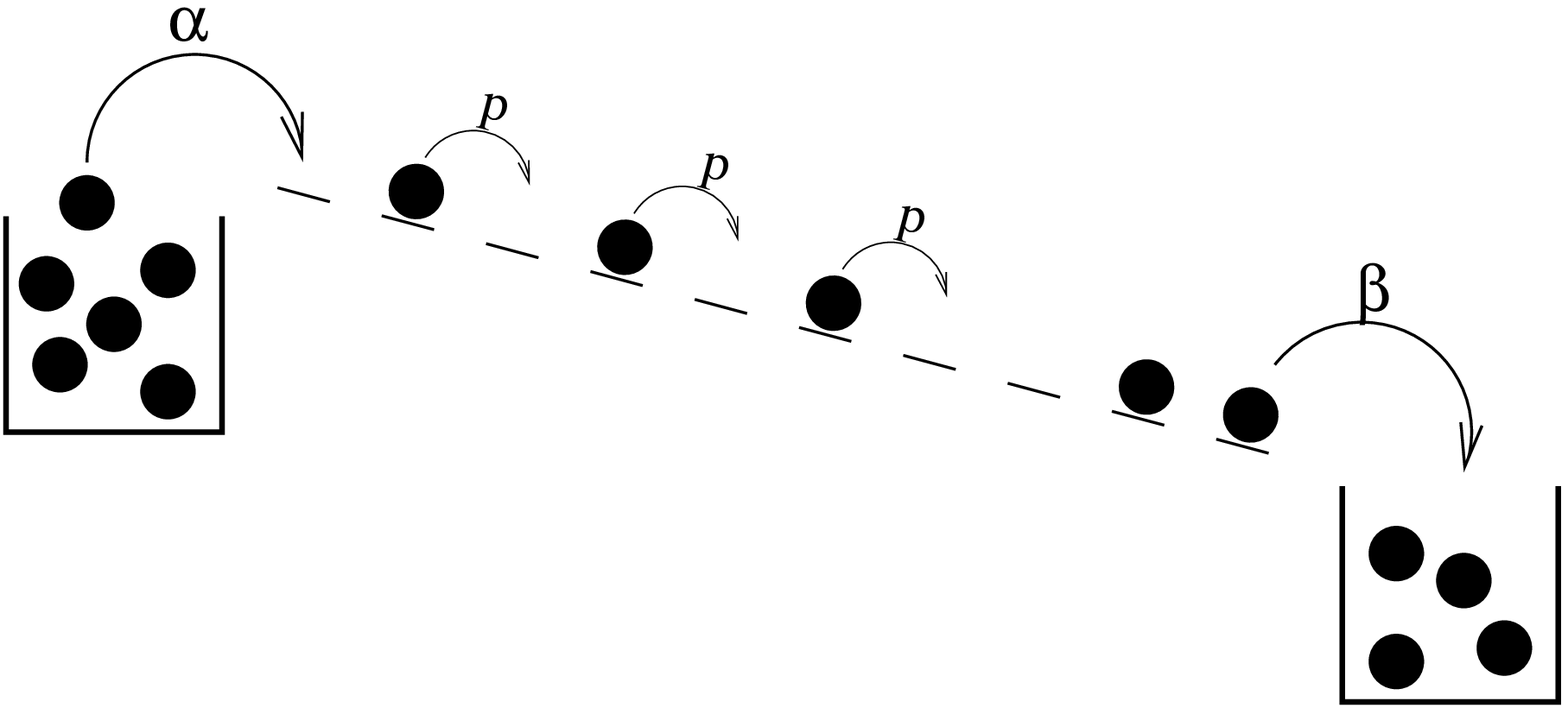}
\end{center}
\caption{Open system with particle input at the left boundary with
rate $\alpha$ and particle output at the right boundary with rate
$\beta$. \label{fig-open} }
\end{figure}

Some of the results can be interpreted in terms of the
well-established theory of boundary-induced phase transitions
\cite{SchadChowNish,Krug1991}. In nonequilibrium systems phase
transitions (in the bulk) can be induced by changing boundary
conditions, generically input and output rates in the case of
transport systems.  A mesoscopic theory has been developed which
allows to derive the phase diagram of an open system (allowing input
and output of particles at the boundaries, see
Figure~\ref{fig-open}) from the fundamental diagram of the periodic
system \cite{Kolomeisky1998}. This theory even makes quantitative
predictions on the basis of an extremal principle \cite{Popkov1999}.

\begin{figure}
\begin{center}
\includegraphics[width=0.35\textwidth]{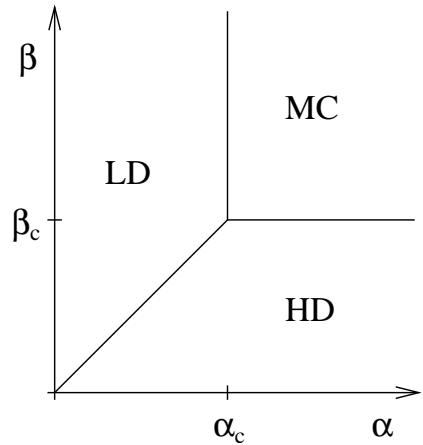}
\end{center}
\caption{\label{fig-ab} Generic form of the phase diagram for an
open system with input rate $\alpha$ and output rate $\beta$. In the
LD phase the current has the form $J=J(\alpha;p)$ and in the HD phase
 $J=J(\beta;p)$. In the MC phase, the current is independent of
$\alpha$ and $\beta$ and corresponds to the maximum of the
fundamental diagram: $J=J_{\rm max}(p)$.}
\end{figure}

The phase diagram as function of the boundary rates $\alpha$ and
$\beta$ has a generic structure. The number of phases observed
depends only on the number of local maxima in the fundamental
diagram. For generic traffic systems it has only one maximum and the
$\alpha$-$\beta$-phase diagram consists of three phases, the
high-density phase (HD), the low-density phase (LD) and the maximum
current phase (MC), see Figure~\ref{fig-ab}. When the supply rate
$\alpha$ of the particles is larger than the removal rate $\beta$
and $\beta < \beta_c$, the particle extraction is the limiting
process resulting in a high density phase where the current is
independent of $\alpha$. When the particles are supplied not too
fast, $\alpha <\alpha_c$ and $\beta > \alpha$, a low density phase
is formed which is limited by particle supply. Here the current is
independent of $\alpha$. There is a discontinuous phase transition
along the line $\alpha = \beta < \alpha_c$. When particles are
supplied and removed sufficiently rapidly, $\alpha > \alpha_c$ and
$\beta >\beta_c$, a continuous phase transition into a
maximum-current phase for which transport is bulk dominated
\cite{Krug1991} occurs where the current is independent of both
$\alpha$ and $\beta$.

The experiments in the corridor geometry can effectively be described
by such a scenario. The input rate $\alpha$ into the corridor is
controlled by the width $b_{\rm entrance}$ of the entrance whereas the
output rate $\beta$ is controlled by the width $b_{\rm exit}$ of the
exit.  Therefore the $\alpha$-$\beta$-phase diagram corresponds in our
case to a $b_{\rm entrance}$-$b_{\rm exit}$-diagram. The width $b_{\rm
  cor}$ of the corridor, on the other hand, controls the maximal
possible bulk flow in the system, given by the maximal flow $J_{\rm max}$
in the fundamental diagram.

For the experiments with $b_{\rm exit}=b_{\rm cor}$ the flow through
the system is not limited by the exit (see Fig.~\ref{fig6}),
corresponding to the case $\beta > \beta_c$. The system is then in the
low-density phase. Here the flow is controlled by the inflow into the
system, i.e.\ effectively by $b_{\rm entrance}$.
At $b_{\rm entrance} \approx 1.45$~m a transition into the maximum
current phase can be observed  (see especially Fig.~\ref{fig6}(a)).
For $b_{\rm entrance}=b_{\rm cor}$ and $b_{\rm exit} < b_{\rm cor}$
the system is in the high-density phase.
The phase diagram will be investigated in more detail in a separate
publication \cite{Zhang-tbp}.

\subsection{Comparison of straight corridor with T-junction}

In this section, we compare the fundamental diagrams for a straight
corridor ($C$) and T-junction ($T$) with channel width $b=2.4$~m.

\begin{figure}
\begin{center}
\includegraphics[scale=1.0]{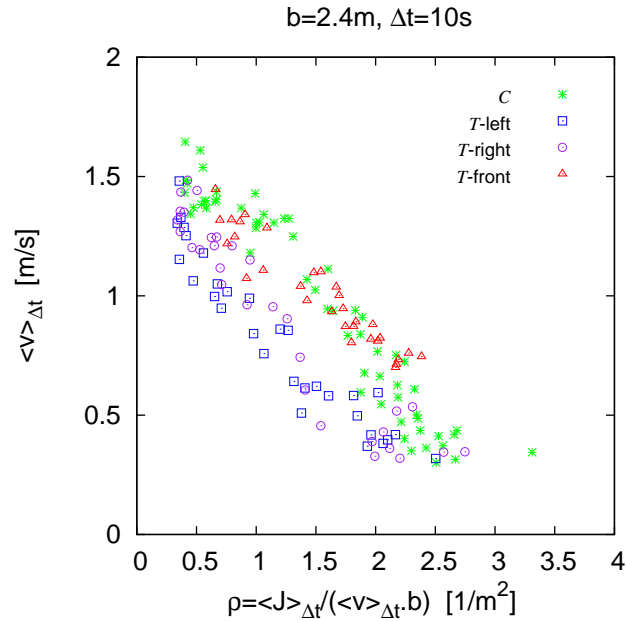}
\caption{\label{fig10} Comparison of the fundamental diagrams
between straight corridor and T-junction}
\end{center}
\end{figure}

Figure \ref{fig10} shows the results obtained from the experiments
$C$ and $T$ using $Method$ $A$.  The data assigned with '$T$-left'
and '$T$-right' are measured in the area before the stream merge
(see Figure \ref{fig1}(c)). The data assigned with '$T$-front' are
measured in the region where the streams already merged. The
comparison shows that the fundamental diagram of the unidirectional
flow agrees with the fundamental diagram of the T-junction in the
'front' part ($T$-front). But for the density region $\rho$ from
0.5~m$^{-2}$ to 2~m$^{-2}$ the velocities at the 'right' and 'left' part
of the T-junction ($T$-left and $T$-right) are significantly lower.
For densities higher than 2~m$^{-2}$ the difference is smaller. The
differences in the fundamental diagram could be interpreted in the
following way. The dynamics in the region after the merging of the
streams is comparable with the unidirectional pedestrian flow in an
open corridor. But in front of the merging the velocities are
significantly lower indicating that the dynamics of the stream
changes due to the change of the geometry and the merging of the
streams.

 \section{Summary}

Series of well-controlled laboratory pedestrian experiments were
performed in straight corridors and T-junction in this study. Up to
350 persons participated in these activities and the whole processes
of the experiment were recorded using two video cameras. The
trajectories of each pedestrian are extracted with high accuracy
from the video recordings automatically using {\em PeTrack}. Four
measurement methods are adopted in this study and their influences
on the fundamental diagram are investigated. It is found that the
results obtained from different methods agree well and the main
differences are the range of the fluctuations and the resolution in
time of the results. The influence of the corridor width on the
results is also investigated. It is shown that fundamental diagrams
for the same type of facility but different widths agree well and
can be unified in one diagram for specific flow. From the comparison
of the fundamental diagrams between straight corridor and
T-junction, it is indicated that the fundamental diagrams for
different facilities are not comparable. The reason for this
difference may be the equilibration between the inflow and outflow
of pedestrians in the corridor. When the outflow and the inflow are not
equal, a transition between low and high densities appears in the
pedestrian flow. This transition can also be observed in the
fundamental diagram.

 \begin{table}[htbp]
   \centering\caption{\label{table2}\centering Parameters for the straight
     corridor experiments } \lineup
 \begin{tabular}{cccccc}
  \toprule
  Experiment index & Name & $b_{\rm entrance}$ [m] & $b_{cor}$ [m] & $b_{\rm exit}$ [m]  & $N$ \\
  \midrule
  1 & $C$-050-180-180 & 0.50 & 1.80 & 1.80 & 61 \\
  2 & $C$-060-180-180 & 0.60 & 1.80 & 1.80 & 66 \\
  3 & $C$-070-180-180 & 0.70 & 1.80 & 1.80 & 111 \\
  4 & $C$-100-180-180 & 1.00 & 1.80 & 1.80 & 121 \\
  5 & $C$-145-180-180 & 1.45 & 1.80 & 1.80 & 175 \\
  6 & $C$-180-180-180 & 1.80 & 1.80 & 1.80 & 220 \\
  7 & $C$-180-180-120 & 1.80 & 1.80 & 1.20 & 170 \\
  8 & $C$-180-180-095 & 1.80 & 1.80 & 0.95 & 159 \\
  9 & $C$-180-180-070 & 1.80 & 1.80 & 0.70 & 148 \\
  10 & $C$-065-240-240 & 0.65 & 2.40 & 2.40 & 70 \\
  11 & $C$-080-240-240 & 0.80 & 2.40 & 2.40 & 118 \\
  12 & $C$-095-240-240 & 0.95 & 2.40 & 2.40 & 108 \\
  13 & $C$-145-240-240 & 1.45 & 2.40 & 2.40 & 155 \\
  14 & $C$-190-240-240 & 1.90 & 2.40 & 2.40 & 218 \\
  15 & $C$-240-240-240 & 2.40 & 2.40 & 2.40 & 246 \\
  16 & $C$-240-240-160 & 2.40 & 2.40 & 1.60 & 276 \\
  17 & $C$-240-240-130 & 2.40 & 2.40 & 1.30 & 247 \\
  18 & $C$-240-240-100 & 2.40 & 2.40 & 1.00 & 254 \\
  19 & $C$-080-300-300 & 0.80 & 3.00 & 3.00 & 119 \\
  20 & $C$-100-300-300 & 1.00 & 3.00 & 3.00 & 100 \\
  21 & $C$-120-300-300 & 1.20 & 3.00 & 3.00 & 163 \\
  22 & $C$-180-300-300 & 1.80 & 3.00 & 3.00 & 208 \\
  23 & $C$-240-300-300 & 2.40 & 3.00 & 3.00 & 296 \\
  24 & $C$-300-300-300 & 3.00 & 3.00 & 3.00 & 349 \\
  25 & $C$-300-300-200 & 3.00 & 3.00 & 2.00 & 351 \\
  26 & $C$-300-300-160 & 3.00 & 3.00 & 1.60 & 349 \\
  27 & $C$-300-300-120 & 3.00 & 3.00 & 1.20 & 348 \\
  28 & $C$-300-300-080 & 3.00 & 3.00 & 0.80 & 270 \\
  \bottomrule
 \end{tabular}
\end{table}

\begin{table}[htbp]
 \centering\caption{\label{table3}\centering Parameters for the T-junction experiments }
 \lineup
 \begin{tabular}{cccccc}
  \toprule
  Experiment index & Name & $b_{\rm cor1}$ [$m$] & $b_{\rm entrance}$ [m] & $b_{\rm cor2}$ [m] & $N$ \\
  \midrule
  1 & $T$-240-050-240 & 2.40 & 0.50 & 2.40 & 134 \\
  2 & $T$-240-060-240 & 2.40 & 0.60 & 2.40 & 132 \\
  3 & $T$-240-080-240 & 2.40 & 0.80 & 2.40 & 228 \\
  4 & $T$-240-100-240 & 2.40 & 1.00 & 2.40 & 208 \\
  5 & $T$-240-120-240 & 2.40 & 1.20 & 2.40 & 305 \\
  6 & $T$-240-150-240 & 2.40 & 1.50 & 2.40 & 305 \\
  7 & $T$-240-240-240 & 2.40 & 2.40 & 2.40 & 302 \\
  \bottomrule
 \end{tabular}
\end{table}

\section*{References}
\bibliographystyle{IEEEtran}
\bibliography{C:/Phdstuff/Literature_Ebook/lit/ped}

\begin{thebibliography}{10}
\providecommand{\url}[1]{#1}
\csname url@rmstyle\endcsname
\providecommand{\newblock}{\relax}
\providecommand{\bibinfo}[2]{#2}
\providecommand\BIBentrySTDinterwordspacing{\spaceskip=0pt\relax}
\providecommand\BIBentryALTinterwordstretchfactor{4}
\providecommand\BIBentryALTinterwordspacing{\spaceskip=\fontdimen2\font plus
\BIBentryALTinterwordstretchfactor\fontdimen3\font minus
  \fontdimen4\font\relax}
\providecommand\BIBforeignlanguage[2]{{%
\expandafter\ifx\csname l@#1\endcsname\relax
\typeout{** WARNING: IEEEtran.bst: No hyphenation pattern has been}%
\typeout{** loaded for the language `#1'. Using the pattern for}%
\typeout{** the default language instead.}%
\else
\language=\csname l@#1\endcsname
\fi
#2}}

\bibitem{Appert-Rolland2009}
C.~Appert-Rolland, F.~Chevoir, P.~Gondret, S.~Lassarre, J.-P. Lebacque, and
  M.~Schreckenberg. \emph{{Traffic and Granular Flow '07}}.\hskip 1em
  plus 0.5em minus 0.4em\relax Springer, Berlin Heidelberg, 2009.

\bibitem{Bandini2010}
\BIBentryALTinterwordspacing
S.~Bandini, S.~Manzoni, H.~Umeo, and G.~Vizzari. \emph{{Cellular Automata: 9th International Conference on 
Cellular Automata for Reseach and Industry}}. ACRI 2010 Ascoli Piceno, Italy, 2010.
\BIBentrySTDinterwordspacing

\bibitem{Klingsch2010}
\BIBentryALTinterwordspacing
W.~Klingsch, C.~Rogsch, A.~Schadschneider, and M.~Schreckenberg.
  \emph{{Pedestrian and Evacuation Dynamics 2008}}.\hskip 1em plus 0.5em minus
  0.4em\relax Springer-Verlag Berlin Heidelberg, 2010.
\BIBentrySTDinterwordspacing

\bibitem{Schadschneider2009c}
\BIBentryALTinterwordspacing
A.~Schadschneider and A.~Seyfried. ``{Empirical Results for Pedestrian Dynamics
  and their Implications for Cellular Automata Models},'' in \emph{{Pedestrian
  Behavior: Data Collection and Applications}}, 1st~ed., ch.~2, pp. 27--43, 2009.
\BIBentrySTDinterwordspacing

\bibitem{Schadschneider2009}
A.~Schadschneider, H.~Kl\"upfel, T.~Kretz, and C.~Rogsch and A.~Seyfried.
  ``{Fundamentals of Pedestrian and Evacuation Dynamics},'' in Bazzan and Kl\"ugl (Eds)
  \emph{{Multi-Agent Systems for Traffic and Transportation Engineering}}, USA, ch.~6, pp. 124--154, 2009. 

\bibitem{SchadChowNish}
A.~Schadschneider, D.~Chowdhury and K.~Nishinari.
``{Stochastic Transport in Complex Systems - From Molecules to Vehicles},''
,Elsevier, 2010

\bibitem{Schadschneider2009a}
A.~Schadschneider, W.~Klingsch, H.~Kluepfel, T.~Kretz, C.~Rogsch, and
  A.~Seyfried. {Evacuation Dynamics: Empirical Results, Modeling and Applications}, in 
  \emph{{Encyclopedia of Complexity and System Science}}. vol.~5, pp. 3142--3176, 2009.

\bibitem{Seyfried2009}
A.~Seyfried, O.~Passon, B.~Steffen, M.~Boltes, T.~Rupprecht, and W.~Klingsch.
  ``{New insights into pedestrian flow through bottlenecks},''
  \emph{Transportation Science}, vol.~43, pp. 395--406, 2009.

\bibitem{Hoogendoorn2005}
S.~P. Hoogendoorn and W.~Daamen, ``{Pedestrian Behavior at Bottlenecks},''
  \emph{Transportation Science}, vol.~39, no.~2, pp. 147--159, 2005.

\bibitem{Kretz2006a}
T.~Kretz, A.~Gr{\"u}nebohm, and M.~Schreckenberg, ``{Experimental study of
  pedestrian flow through a bottleneck},'' \emph{J. Stat. Mech.}, vol.~10, p.
  P10014, 2006.

\bibitem{Kretz2006}
T.~Kretz, A.~Gr{\"u}nebohm, M.~Kaufman, F.~Mazur, and M.~Schreckenberg,
  ``{Experimental study of pedestrian counterflow in a corridor},'' \emph{J.
  Stat. Mech.}, vol.~10, p. P10001, 2006.

\bibitem{Moussaid2009}
M.~Moussaid, D.~Helbing, S.~Garnier, A.~Johansson, M.~Combe, and
  G.~Theraulaz, ``{Experimental study of the behavioural mechanisms underlying
  self-organization in human crowds.}'' \emph{Proc. R. Soc. B}, vol. 276, no.
  1668, pp. 2755--2762, 2009.

\bibitem{Liu2009}
X.~Liu, W.~Song, and J.~Zhang, ``{Extraction and quantitative analysis of
  microscopic evacuation characteristics based on digital image processing},''
  \emph{Physica A: Statistical Mechanics and its Applications}, vol. 388,
  no.~13, pp. 2717--2726, 2009.

\bibitem{Johansson2009a}
A.~Johansson, ``{Constant-net-time headway as a key mechanism behind pedestrian
  flow dynamics},'' \emph{Phys. Rev. E}, vol.~80, 026120, 2009.

\bibitem{Johansson2008}
A.~Johansson and D.~Helbing, ``{From crowd dynamics to crowd safety: a
  video-based analysis},'' \emph{Advances in Complex Systems (ACS)}, vol.~4,
  no.~4, pp. 497--527, 2008.

\bibitem{Young1999}
S.~B. Young, ``{Evaluation of Pedestrian Walking Speeds in Airport
  Terminals},'' \emph{Transportation Research Record}, vol. 1674, pp. 20--26,
  1999.

\bibitem{Fruin1971}
J.~J. Fruin. \emph{{Pedestrian Planning and Design}}.\hskip 1em plus 0.5em
  minus 0.4em\relax Elevator World, New York, 1971.

\bibitem{Predtechenskii1978}
V.~M. Predtechenskii and A.~I. Milinskii. \emph{{Planning for Foot Traffic Flow
  in Buildings}}. Amerind Publishing,
  New Delhi, 1978, translation of: Proekttirovanie Zhdanii s Uchetom
  Organizatsii Dvizheniya Lyuddskikh Potokov, Stroiizdat Publishers, Moscow,
  1969.

\bibitem{Weidmann1993}
U.~Weidmann. ``{Transporttechnik der Fussg\"anger},'' Schriftenreihe des Institut f\"ur
Verkehrsplanung,Transporttechnik, Strassen- und Eisenbahnbau, ETH Z\"urich, 1993), Vol. 90.

\bibitem{Helbing2007}
D.~Helbing, A.~Johansson, and H.~Z. Al-Abideen, ``{Dynamics of Crowd Disasters:
  An Empirical Study},'' \emph{Physical Review E}, vol.~75, p. 046109, 2007.

\bibitem{Seyfried2010}
\BIBentryALTinterwordspacing
A.~Seyfried, M.~Boltes, J.~K\"ahler, W.~Klingsch, A.~Portz, T.~Rupprecht,
  A.~Schadschneider, B.~Steffen, and A.~Winkens. ``{Enhanced empirical data for
  the fundamental diagram and the flow through bottlenecks},'' in
  \emph{Pedestrian and Evacuation Dynamics 2008}.\hskip 1em plus 0.5em minus
  0.4em\relax Springer-Verlag Berlin Heidelberg, pp. 145--156, 2010.
\BIBentrySTDinterwordspacing

\bibitem{Chattaraj2009}
U.~Chattaraj, A.~Seyfried, and P.~Chakroborty. ``{Comparison of Pedestrian
  Fundamental Diagram Across Cultures},'' \emph{Advances in Complex Systems
  (ACS)}, vol.~12, no.~3, pp. 393--405, 2009.

\bibitem{Navin1969}
F.~D. Navin and R.~J. Wheeler. ``{Pedestrian flow characteristics},''
  \emph{Traffic Engineering}, vol.~39, pp. 31--36, 1969.

\bibitem{Pushkarev1975}
B.~Pushkarev and J.~M. Zupan. ``{Capacity of Walkways},'' \emph{Transportation
  Research Record}, vol. 538, pp. 1--15, 1975.

\bibitem{Hankin1958}
B.~D. Hankin and R.~A. Wright. ``{Passenger Flow in Subways},''
  \emph{Operational Research Quarterly}, vol.~9, pp. 81--88, 1958.
  
\bibitem{Hermes}
http://www.fz-juelich.de/jsc/hermes  

\bibitem{Boltes2010}
M.~Boltes, A.~Seyfried, B.~Steffen, and A.~Schadschneider. ``{Automatic
  Extraction of Pedestrian Trajectories from Video Recordings},'' in
  \emph{Pedestrian and Evacuation Dynamics 2008}.\hskip 1em plus 0.5em minus
  0.4em\relax Springer-Verlag Berlin Heidelberg, pp. 43--54, 2010.

\bibitem{Leutzbach1988}
W.~Leutzbach. \emph{{Introduction to the Theory of Traffic Flow}}.\hskip 1em
  plus 0.5em minus 0.4em\relax Springer, 1988.

\bibitem{Kerner2004}
B.~S. Kerner. \emph{{The Physics Of Traffic: Empirical Freeway Pattern
  Features, Engineering Applications, and Theory}}, 1st~ed., J.~A.~S. Kelso,
  Ed.\hskip 1em plus 0.5em minus 0.4em\relax Springer, 2004.

\bibitem{Steffen2010a}
B.~Steffen and A.~Seyfried. ``{Methods for measuring pedestrian density, flow,
  speed and direction with minimal scatter},'' \emph{Physica A}, vol. 389,
  no.~9, pp. 1902--1910, 2010.

\bibitem{Voronoi1908}
G.~M. Voronoi. ``{Nouvelles applications des param\`etres continus \`a la
  th\'eorie des formes quadratiques},'' \emph{Journal f\"ur die reine und
  angewandte Mathematik}, vol. 133, pp. 198--287, 1908.

\bibitem{Krug1991}
J.~Krug. ``{Boundary-Induced Phase Transition in Driven Diffusive Systems},''
  \emph{Phys. Rev. Lett.}, vol.~67, no.~14, pp. 1882--1885, 1991.

\bibitem{Kolomeisky1998}
A.~Kolomeisky, G.~Sch\"utz, E.~Kolomeisky, and J.~Straley. ``Phase diagram of
  one-dimensional driven lattice gases with open boundaries,'' \emph{J. Phys.
  A}, vol.~31, pp. 6911--6919, 1998.

\bibitem{Popkov1999}
V.~Popkov and G.~Sch\"utz. ``{Steady-state selection in driven diffusive
  systems with open boundaries},'' \emph{Europhys. Lett.}, vol.~48, no.~3, pp.
  257--263, 1999.

\bibitem{Zhang-tbp}
J.~Zhang, A.~Schadschneider and A.~Seyfried. ``{in preparation},''

\end{thebibliography}
\end{document}